\documentclass[journal=jpccck,manuscript=article,layout=twocolumn]{achemso}

\setkeys{acs}{abbreviations=true,articletitle=true,chaptertitle=true,doi=true,email=true,etalmode=truncate,keywords=true,maxauthors=100,super=true}
\setcitestyle{super,open={},close={}}

\usepackage[T1]{fontenc}
\usepackage[utf8]{inputenc}
\usepackage[english]{babel}
\usepackage{booktabs}
\usepackage[format=plain,singlelinecheck=false,font={small,bf},labelfont=bf,labelsep=space]{caption}
\usepackage[version=4]{mhchem}
\usepackage{mathpazo}
\usepackage{siunitx}
\usepackage[none]{hyphenat}
\usepackage{microtype}
\usepackage{mathtools}
\usepackage{soul}
\usepackage{tabularx}
\usepackage{multirow}
\usepackage{graphicx}
\usepackage{float}
\usepackage{placeins}
\usepackage{xcolor}
\usepackage{indentfirst}
\usepackage{longtable}
\usepackage{multicol}
\usepackage{subfig}
\usepackage[breaklinks,colorlinks=true,allcolors=blue]{hyperref}
\hypersetup{citecolor=blue,urlcolor=blue}
\providecommand{\doi}[1]{\href{https://doi.org/#1}{\nolinkurl{#1}}}
\DeclareUnicodeCharacter{2212}{\ensuremath{-}}
\newcommand{\cNb}{\textit{c}-\ce{Nb}}
\SectionsOn
\SectionNumbersOn
\AbstractOn

\vfuzz \hfuzz
\author{Willian O. Santos}
\email{willnp_96@hotmail.com}
\affiliation[PPGF-UFS]{Graduate Program in Physics, Federal University of Sergipe (UFS), 49107-230, Aracaju, SE, Brazil}

\author{José A. S. Laranjeira}
\affiliation[Unesp]{Post-Graduate Program in Materials Science and Technology (POSMAT), São Paulo State University (UNESP), School of Sciences, 17033-360, Bauru, SP, Brazil}

\author{Kleuton A. L. Lima}
\affiliation[UNICAMP]{Department of Applied Physics and Center for Computational Engineering and Sciences, State University of Campinas (UNICAMP), 13083-859, Campinas, SP, Brazil}

\author{Douglas S. Galvão}
\affiliation[UNICAMP]{Department of Applied Physics and Center for Computational Engineering and Sciences, State University of Campinas (UNICAMP), 13083-859, Campinas, SP, Brazil}

\author{David L. Azevedo}
\affiliation[IF-UnB]{Dynamics and Ab Initio Group, Institute of Physics, University of Brasília, 70910-900, Brasília, DF, Brazil}

\title{Corrugated Niobene: Stability, Mechanical and Electronic Properties from First-Principles Calculations}

\date{\today}

\begin{document}

\begin{abstract}
The growing diversity of two-dimensional (2D) materials has expanded opportunities in electronics, energy technologies, catalysis, and optical applications. Within this landscape, understanding how corrugated architectures accommodate metallic bonding and influence the stability and properties of monoelemental niobium sheets remains an open research direction. Here, we propose a corrugated niobene nanosheet, denoted \textit{c}-\ce{Nb}, and investigate its structural, energetic, vibrational, mechanical, electronic, and optical properties using density functional theory simulations. The optimized structure adopts a centrosymmetric $P\bar{3}m1$ symmetry, with six atoms per primitive cell and a geometrical thickness of \SI{4.279}{\angstrom}. Its cohesive energy of $-6.03~\mathrm{eV\,atom^{-1}}$ indicates strong binding relative to isolated atoms, although bulk BCC \ce{Nb} remains energetically favored by $2.49~\mathrm{eV\,atom^{-1}}$. The absence of imaginary phonon frequencies along the investigated high-symmetry path and the positive-definite in-plane stiffness matrix support dynamical and mechanical structural stability. Despite its pronounced corrugation, the nanosheet exhibits isotropic in-plane elasticity, with Young's and shear moduli of \SI{43.66}{\newton\per\metre} and \SI{13.66}{\newton\per\metre}, respectively, and a Poisson's ratio of 0.598. Electronic states near the Fermi level are dominated by \ce{Nb} $4d$ orbitals, establishing multiband metallicity. These findings identify a corrugated structural candidate that combines strong cohesion, elastic compliance, and metallic electronic behavior, extending the structural diversity of monoelemental 2D metals and providing a basis for further exploration of niobium nanosheets in electronic and ultraviolet-responsive systems.
\end{abstract}
\vspace{5mm}

\section{Introduction}\label{sec:introduction}

The experimental realization of graphene established atomically thin materials as a platform for studying how reduced dimensionality modifies solid-state properties~\cite {doi:10.1126/science.1102896}. This has recently been complemented by the discovery of novel carbon allotropes\cite{desyatkin2022scalable}, such as planar non-graphitic sp² carbon phases produced via low-temperature reactions of $\gamma$-graphyne precursors\cite{aliev2025planar}. This development stimulated interest in monoelemental two-dimensional (2D) materials\cite{yang2025graphene, rana2026group, mishra2025ultrahigh}, whose chemical simplicity accommodates considerable structural and electronic diversity. Their atomic arrangements extend beyond planar sheets to buckled and puckered geometries, while their electronic behavior ranges from semiconductors to metallic, or even superconductors. These materials therefore provide opportunities to examine how elemental composition and local coordination govern physical properties at the 2D limit, as illustrated by the contrasting structures and electronic responses of borophene, silicene, and phosphorene~\cite{xu2018computational, kumar2025recent, shan2023recent}.

Experimental advances have established several representatives of this family. Researchers have grown borophene on silver surfaces and incorporated atomically thin silicon and few-layer black phosphorus into field-effect transistors~\cite{doi:10.1126/science.aad1080,Tao2015,Li2014}. These developments demonstrate that monoelemental sheets can support distinct bonding environments and device functionalities. More recently, reports of molybdenene synthesis from \ce{MoS2} and goldene preparation by selective etching of a nanolaminated precursor have extended this research toward elemental metals~\cite{Sahu2023,Kashiwaya2024}. Such routes broaden the experimental approaches available for producing ultrathin metallic sheets from precursors without relying on the exfoliation of a layered elemental parent crystal.

The applications of monoelemental 2D materials reflect this diversity of electronic and structural properties\cite{bafekry2022monoelemental,gusarov2024comprehensive, xu2025janus, yu2025novel}. Silicene and few-layer black phosphorus have demonstrated their relevance to electronic devices through gate-controlled charge transport~\cite{Tao2015,Li2014}, while molybdenene has been investigated for molecular sensing using surface-enhanced Raman spectroscopy~\cite{Sahu2023}. For metallic sheets, theoretical studies also point to opportunities in energy storage and optical applications. Zulfiqar et al. identified a stable silverene candidate and predicted quantum-capacitance characteristics favorable for supercapacitor electrodes~\cite{ZULFIQAR2025107452}. In a complementary study, dos Santos et al. examined goldene, silverene, and copperene, reporting metallic electronic structures together with distinct elastic and optical responses~\cite{10.1021/acsomega.5c01823}.

Goldene has also attracted growing interest as a promising surface for molecular adsorption and for studying chemical reactions relevant to energy conversion and storage applications, as well as for developing new catalytic materials\cite{Sheremetyeva_2025, picello2025single, rahimi2025trifunctional}. Kumar et al. investigated dopant-controlled \ce{CO2} activation and found that \ce{Be}-doped goldene promotes chemisorption and molecular bending, whereas alkali-metal-doped sheets retain linear molecular geometry~\cite{KUMAR2025164509}. Hydrogen-related studies have addressed distinct adsorption regimes: Sheremetyeva and Meunier examined atomic hydrogen at different coverages and its effects on the electronic and vibrational properties of goldene~\cite{Sheremetyeva_2025}, while Iyakutti et al. predicted nondissociative \ce{H2} adsorption with an adsorption energy of \SI{-0.175}{\electronvolt}~\cite{IYAKUTTI2025e01132}. In the context of lithium--sulfur batteries, Martins et al. reported strong lithium polysulfide anchoring and favorable conversion energetics on goldene, associated with interfacial charge redistribution and hybridization between \ce{Au} $d$ and \ce{S} $p$ states~\cite{10.1039/d6ta00772d}. Together, these theoretical studies illustrate how surface chemistry and electronic structure can be exploited in monoelemental 2D metals for molecular activation, functionalization, and electrochemical applications.

Beyond surface functionality, identifying stable atomic arrangements remains central to expanding this material family. Demirci et al. predicted stable \ce{Mo} bilayers assembled from zigzag chains and analogous structures for \ce{Cr} and \ce{W}, demonstrating that multiple atomic planes can support low-dimensional transition-metal structures with distinct electronic and elastic properties~\cite{10.1021/acs.jpcc.4c03419}. In parallel, niobium-containing transition-metal compounds and nanostructures have recently attracted significant attention for energy conversion and optoelectronic applications, particularly because of their favorable light-harvesting characteristics, catalytic activity, and stability in photovoltaic and solar-driven devices~\cite{PONNALAGAR2023112046, marques2026first}. These findings motivate exploring corrugated architectures in other transition metals, where out-of-plane coordination provides an additional degree of freedom to accommodate metal--metal bonding.

Niobium offers a path to extend this investigation to 2D transition-metal structures with partially occupied $4d$ states. In an atomically thin \ce{Nb} sheet, changes in coordination and out-of-plane geometry can modify orbital overlap, linking the atomic arrangement to bonding and electronic behavior. A corrugated configuration is therefore a relevant candidate for examining how finite thickness within a single sheet accommodates \ce{Nb}\textendash\ce{Nb} interactions. The central questions are whether such an arrangement is stable against structural distortions and how its geometry influences the mechanical and electronic properties of the resulting material.

In this context, we propose and characterize a corrugated niobene nanosheet, denoted \cNb{}, using first-principles density functional theory simulations. The analysis encompasses its optimized geometry and bonding, energetic, dynamical, and mechanical stability, electronic band structure, its corresponding density of states, and simulated optical absorption. The results link the corrugated architecture to elastic and electronic responses, expanding the structural diversity of monoelemental 2D metals and providing a basis for further exploration of atomically thin niobium.

\section{Methodology}
\label{sec:methods}

First-principles calculations were performed using Quantum ESPRESSO~\cite{Giannozzi_2009} within the framework of density functional theory (DFT)~\cite{hohenberg1964inhomogeneous,kohn1965self}. Exchange and correlation were described using the Perdew--Burke--Ernzerhof (PBE) functional within the generalized gradient approximation (GGA)~\cite{PhysRevLett.77.3865}. Core--valence interactions were treated using the projector augmented-wave (PAW) method~\cite{PhysRevB.50.17953}.

The Kohn--Sham wavefunctions were expanded in a plane-wave basis with a kinetic-energy cutoff of $60~\mathrm{Ry}$, while a cutoff of $480~\mathrm{Ry}$ was adopted for the charge density and potential. Brillouin-zone integrations employed a $24 \times 24 \times 1$ Monkhorst--Pack $k$-point mesh~\cite{monkhorst1976special} for both structural optimization and electronic-property calculations. Electronic occupations were treated using Gaussian smearing with a width of $0.04~\mathrm{Ry}$. The self-consistent electronic calculations employed Davidson diagonalization and plain density mixing with a mixing factor of $0.7$. Electronic convergence was achieved when the estimated energy error fell below $10^{-8}~\mathrm{Ry}$.

Atomic positions were optimized using the Broyden--Fletcher--Goldfarb--Shanno (BFGS) algorithm while keeping the lattice vectors fixed. The force convergence threshold was set to $10^{-4}~\mathrm{Ry\,bohr^{-1}}$.

\section{Results and discussion}
\label{sec:results-and-discussion}

\subsection{Structural properties}

\begin{figure*}[htbp]
\centering
\includegraphics[width=\linewidth]{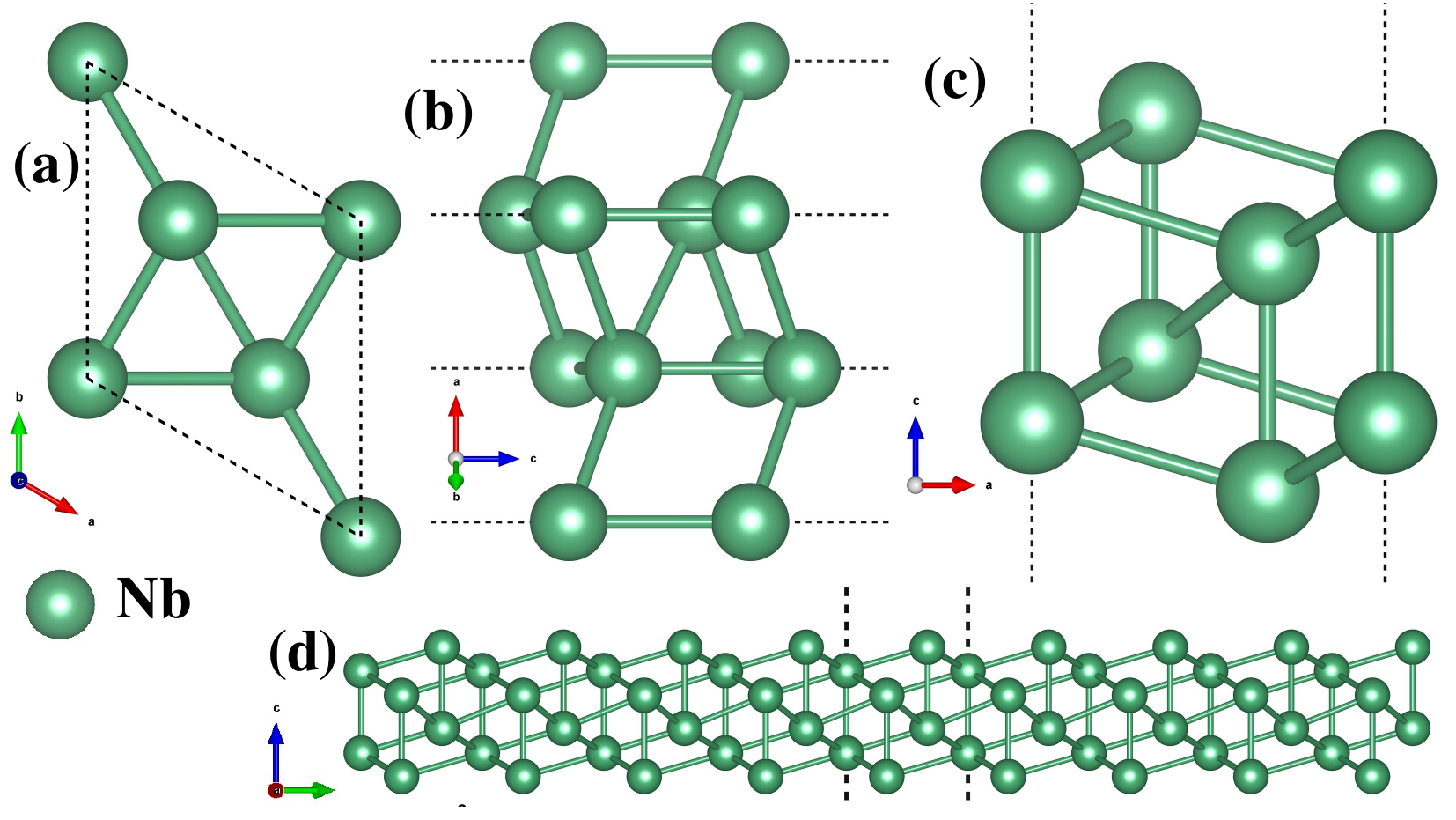}
\caption{Optimized atomic structure of \cNb{}.
(a) Top view of the unit cell, highlighting the atomic arrangement and in-plane connectivity of Nb atoms.
(b,c) Different side views of the unit cell, illustrating the out-of-plane atomic arrangement and the corrugated geometry of the monolayer.
(d) Extended view of the periodic structure, highlighting the connectivity and structural periodicity of the Nb framework.}
\label{structure}
\end{figure*}

Fig.~\ref {structure} shows the equilibrium geometry of the corrugated niobene nanosheet (\cNb{}). The structure has a hexagonal in-plane lattice and belongs to the centrosymmetric space group $P\bar{3}m1$ (No.~164) in the trigonal crystal system in hexagonal setting. The optimized in-plane lattice parameters are $a=b=\SI{4.631}{\angstrom}$, with $\alpha=\beta=90^{\circ}$ and $\gamma=120^{\circ}$.

The asymmetric unit cell comprises three crystallographically inequivalent \ce{Nb} sites, denoted Nb1, Nb3, and Nb5. Table~\ref {parameters} summarizes their fractional coordinates, Wyckoff positions, and site symmetries. Nb1 and Nb5 occupy $2d$ positions, whereas Nb3 occupies a $2c$ position; all three sites have $3m$ site symmetry. The space-group symmetry operations generate two atoms from each inequivalent site, yielding a six-atom primitive unit cell.

\begin{table}[htbp]
\centering
\caption{Fractional coordinates, Wyckoff positions, and site symmetries of the inequivalent \ce{Nb} atoms in the optimized \cNb{} primitive cell.}
\label{parameters}
\renewcommand{\arraystretch}{1.15}
\begin{tabular}{lcccc}
\toprule
Atom & Site (symmetry) & $x$ & $y$ & $z$ \\
\midrule
Nb1 & $2d\;(3m)$ & $2/3$ & $1/3$ & 0.41034 \\
Nb3 & $2c\;(3m)$ & 0 & 0 & 0.44291 \\
Nb5 & $2d\;(3m)$ & $1/3$ & $2/3$ & 0.47715 \\
\bottomrule
\end{tabular}
\end{table}

The six atomic planes occur at fractional heights $z=0.41034$, $0.44291$, $0.47715$, $0.52285$, $0.55709$, and $0.58966$. They are symmetrically distributed about $z=0.50000$, consistent with inversion symmetry. The separation between the outermost planes ($h$) is \SI{4.279}{\angstrom}, providing a geometrical measure of the nanosheet thickness. These planes form a single connected corrugated framework. Considering \ce{Nb-Nb} contacts shorter than \SI{3.0}{\angstrom}, the interatomic distances range from \SI{2.685}{\angstrom} to \SI{2.887}{\angstrom}. For comparison, bulk BCC \ce{Nb} optimized at the same level of theory exhibits a nearest-neighbor distance of \SI{2.876}{\angstrom}, in close agreement with the experimental value of \SI{2.882}{\angstrom}~\cite{10.1021/ja051508a}. The nanosheet therefore accommodates both shorter contacts and separations close to the bulk values.

\subsection{Energetics and dynamical structural stability}

The cohesive energy per atom, $E_{\mathrm{coh}}$, was evaluated relative to isolated \ce{Nb} atoms~\cite{santos20241t,santos2025dft} using the equation:

\begin{equation}
E_{\mathrm{coh}}=
\frac{E_{\mathrm{sheet}}-N E_{\mathrm{atom}}^{\mathrm{Nb}}}{N},
\label{eq:cohesive_energy}
\end{equation}

where $E_{\mathrm{sheet}}$ is the total energy of the primitive cell, $N=6$ is its number of atoms, and $E_{\mathrm{atom}}^{\mathrm{Nb}}$ is the energy of an isolated \ce{Nb} atom. With this sign convention, the calculated cohesive energy of $-6.03~\mathrm{eV\,atom^{-1}}$ indicates strong binding relative to isolated atoms. For comparison, bulk BCC \ce{Nb}, optimized at the same level of theory, exhibits a cohesive energy of $-8.52~\mathrm{eV\,atom^{-1}}$. Thus, the corrugated nanosheet lies $2.49~\mathrm{eV\,atom^{-1}}$ above the bulk reference, showing that BCC \ce{Nb} remains energetically favored despite the strong cohesion of the 2D structure.

The reported cohesive energies of goldene, silverene, and copperene are approximately $-2.9$, $-2.3$, and $-3.1~\mathrm{eV\,atom^{-1}}$, respectively~\cite{10.1021/acsomega.5c01823}. The more negative value obtained for \cNb{} indicates stronger average cohesion within this comparison. Since these energies are normalized per atom, this difference is not simply a consequence of the six-atom primitive cell of \cNb{}.

\begin{figure}[htbp]
\centering
\includegraphics[width=0.9\linewidth]{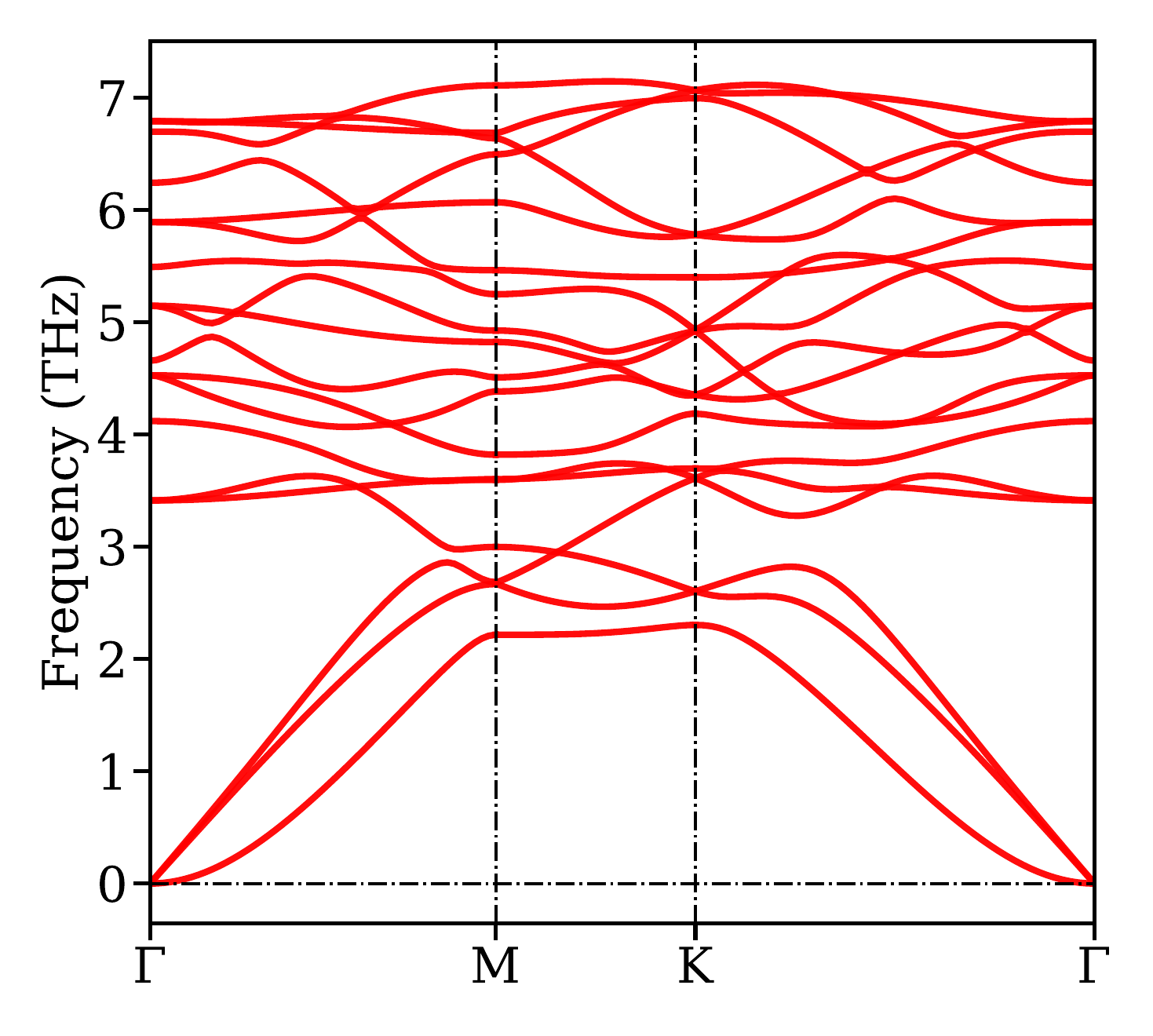}
\caption{Phonon dispersion of \cNb{} along the high-symmetry path $\Gamma$--M--K--$\Gamma$ in the first Brillouin zone.}
\label{fig:phonons}
\end{figure}

The phonon dispersion in Fig.~\ref{fig:phonons} contains 18 branches, comprising three acoustic and 15 optical modes, as expected for a six-atom primitive cell. No imaginary frequencies are observed along the sampled $\Gamma$--M--K--$\Gamma$ path, supporting the dynamical stability of the optimized structure within the harmonic approximation. Near $\Gamma$, the in-plane acoustic branches are approximately linear, while the out-of-plane flexural branch (ZA) exhibits pronounced curvature.

The acoustic branches extend to approximately \SI{3.0}{\tera\hertz}, with a small separation from the lowest optical modes in the calculated dispersion. Several optical branches between approximately \SI{3.5}{\tera\hertz} and \SI{7.2}{\tera\hertz} are flatter than the acoustic branches, indicating smaller group velocities. Because the system is monoelemental, the different frequency ranges arise from collective displacement patterns, local force constants, and the polarizations of the six \ce{Nb} atoms, rather than from differences in atomic mass. Crossings and near-degeneracies around M and K further reflect the primitive cell's vibrational complexity.

\begin{figure}[htbp]
\centering
\includegraphics[width=0.9\linewidth]{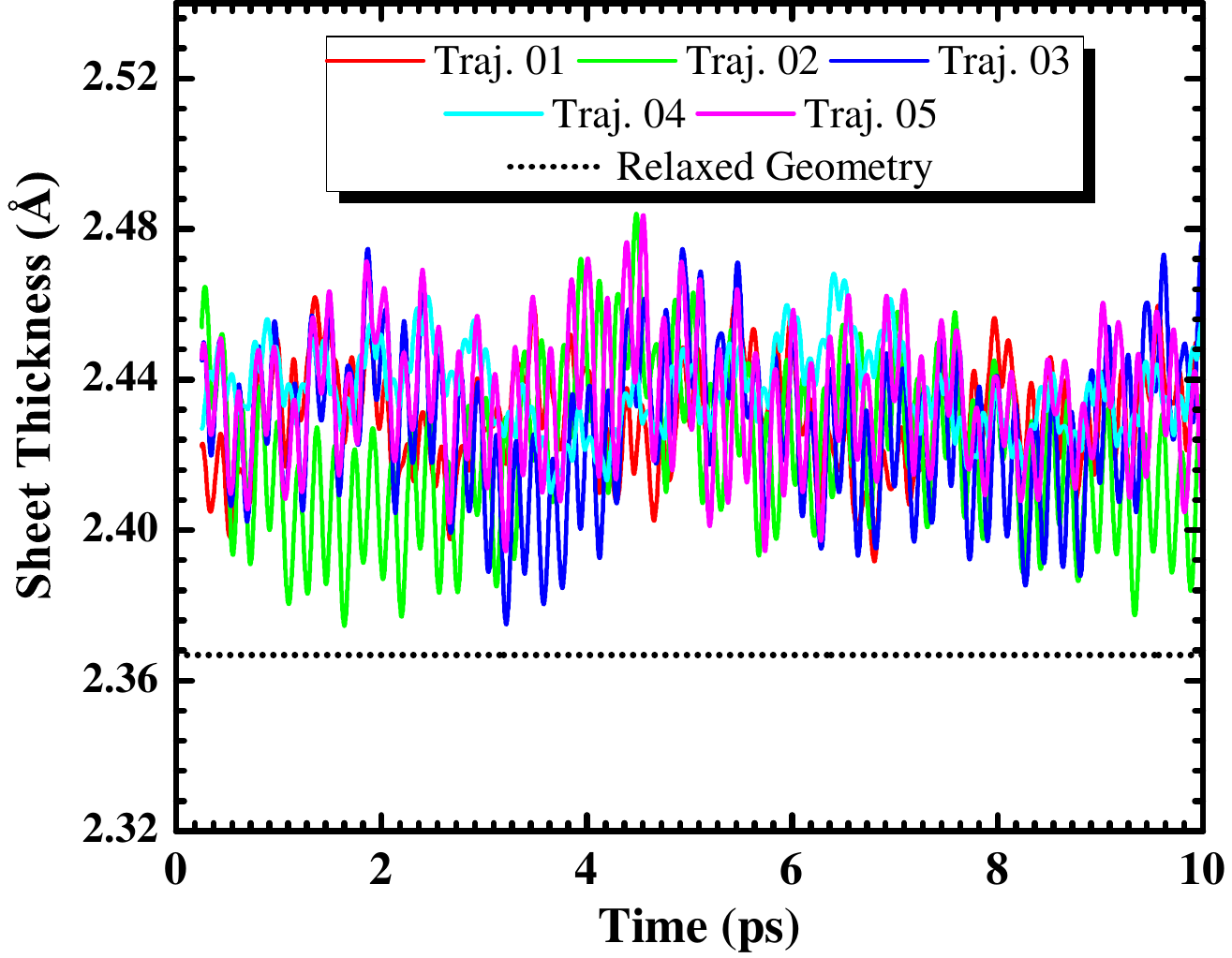}
\caption{Time evolution of the sheet thickness of \cNb{} during the AIMD simulations over a 10~ps interval. The colored curves represent the five trajectories, while the horizontal dotted line indicates the corresponding value for the relaxed geometry.}
\label{fig:sheet_thickness}
\end{figure}

To complement the harmonic phonon analysis, we further examined the structural behavior of \cNb{} at finite temperature by monitoring the sheet thickness during ab initio molecular dynamics (AIMD) simulations. As shown in Fig.~\ref{fig:sheet_thickness}, all five trajectories exhibit limited fluctuations throughout the 10~ps simulation interval. The instantaneous values remain predominantly within approximately $2.39$--$2.48$~\AA, with most fluctuations occurring around $2.40$--$2.46$~\AA, compared with approximately $2.37$~\AA\ for the relaxed geometry.

The monitored sheet thickness shows no progressive increase or decrease throughout the simulations, and the five trajectories behave comparably. Finite-temperature fluctuations therefore produce moderate structural distortions, with no evidence of systematic evolution of this geometrical parameter over the simulated time scale.

Together with the absence of imaginary phonon frequencies, these AIMD results provide additional evidence for the structural robustness of the corrugated \cNb{} framework under the investigated simulation conditions.

\subsection{Electronic properties}

\begin{figure}[htbp]
\centering
\includegraphics[width=\linewidth]{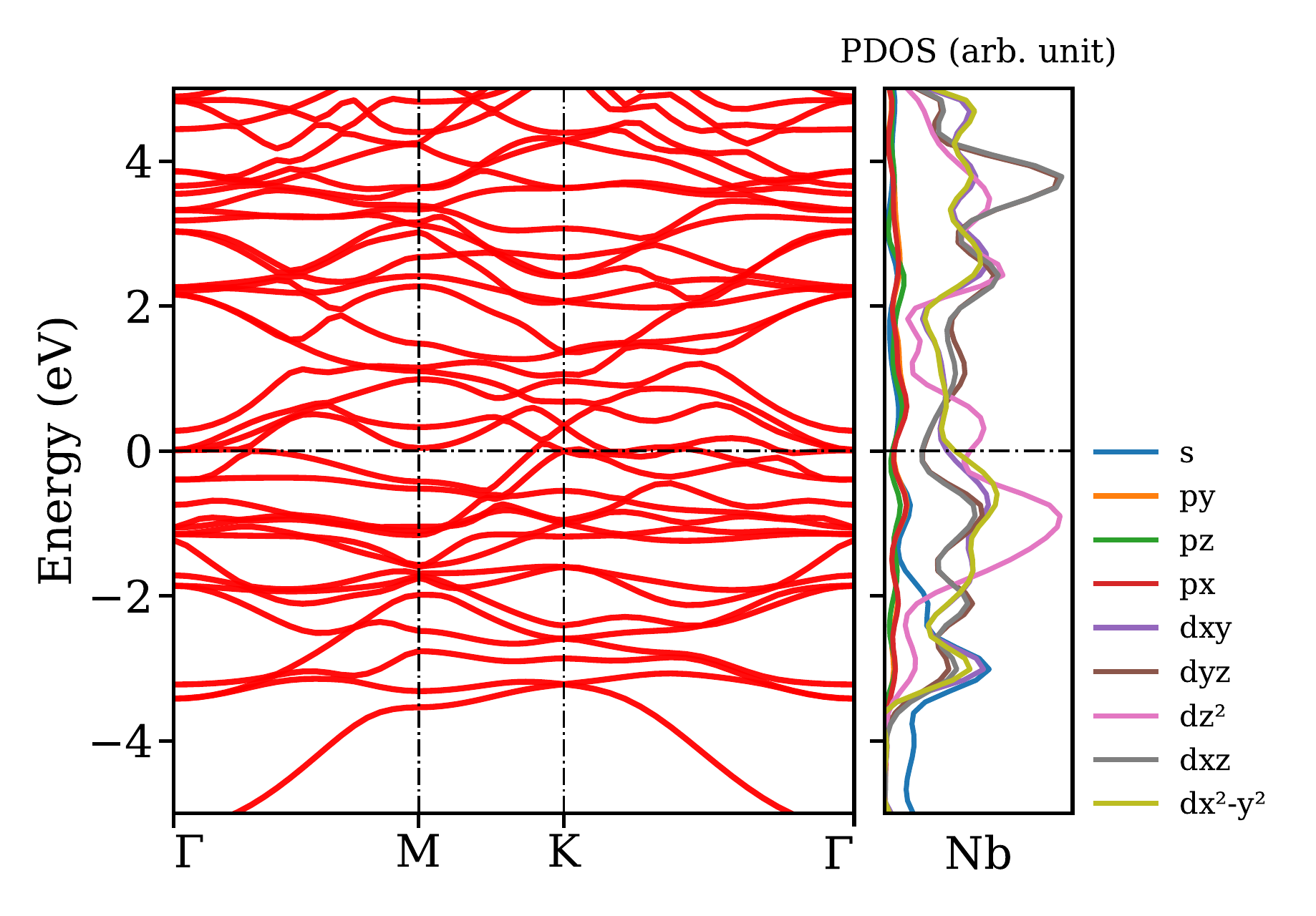}
\caption{PBE electronic band structure of \cNb{} along $\Gamma$--M--K--$\Gamma$ (left) and corresponding orbital-projected density of states (right). The Fermi level is set to \SI{0}{\electronvolt}. Multiple band crossings and a finite \ce{Nb} $4d$ contribution at $E_{\mathrm{F}}$ indicate multiband metallicity.}
\label{fig:bands}
\end{figure}

We have calculated the PBE electronic band structure along the $\Gamma$--M--K--$\Gamma$ path and analyzed it alongside the orbital-projected density of states (PDOS). As shown in Fig.~\ref{fig:bands}, multiple bands cross the Fermi level, $E_{\mathrm{F}}$, and the PDOS remains finite at this energy, establishing the multiband metallic character of \cNb{}.

The low-energy region contains dispersive and intersecting bands. Locally linear segments near K and along M--K and K--$\Gamma$ occur within a dense set of electronic states and do not, by themselves, establish an isolated Dirac cone. The states near $E_{\mathrm{F}}$ are predominantly derived from the \ce{Nb} $4d$ manifold, with contributions from $d_{xy}$, $d_{yz}$, $d_{z^2}$, $d_{xz}$, and $d_{x^2-y^2}$ orbitals. The \ce{Nb} $s$ and $p$ components are comparatively small in this region. The participation and dispersion of the $4d$ states are consistent with orbital hybridization and substantial overlap between neighboring \ce{Nb} atoms.

\subsection{Optical properties}

\begin{figure}
    \centering
    \includegraphics[width=\linewidth]{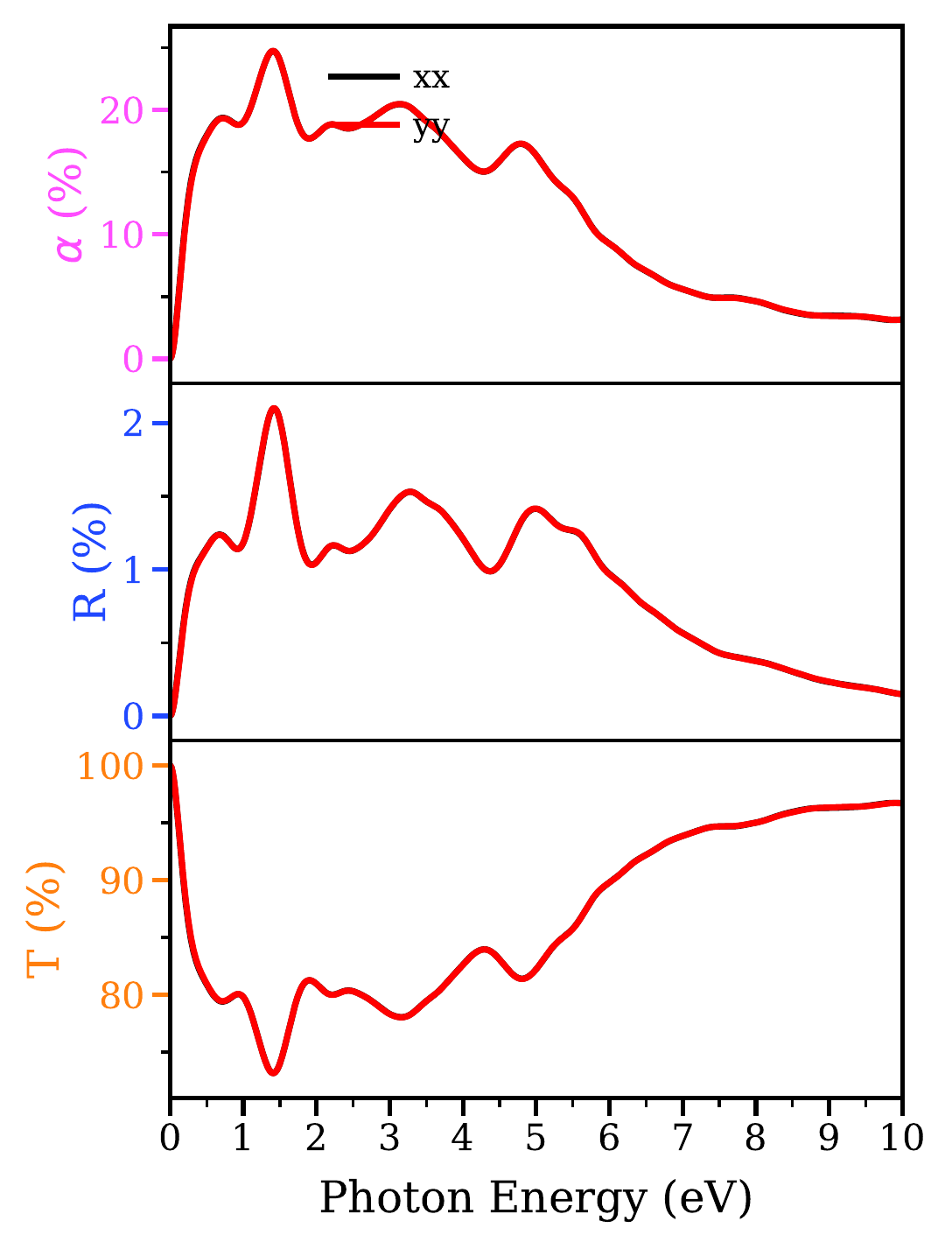}
    \caption{Optical response of \cNb{} calculated at the PBE level: absorptance (top), reflectance (middle), and transmittance (bottom) as functions of photon energy. The $xx$ and $yy$ labels denote orthogonal in-plane Cartesian polarizations, whose spectra overlap owing to the crystal symmetry. All quantities are expressed as percentages of the incident optical power.}
    \label{fig:optical}
\end{figure}

The optical response of \cNb{} was evaluated at the PBE level. Fig.~\ref{fig:optical} presents the absorptance ($\alpha$), reflectance ($R$), and transmittance ($T$) as functions of photon energy. These quantities represent the fractions of incident optical power absorbed, reflected, and transmitted by the nanosheet, respectively, and are expressed as percentages. For a two-dimensional material, this representation describes the optical response of the sheet without assigning an effective material thickness, which would be required to interpret a bulk-like absorption coefficient in inverse-length units.

The $xx$ and $yy$ labels denote orthogonal Cartesian polarizations within the nanosheet plane, rather than the nonorthogonal primitive lattice vectors $\mathbf{a}$ and $\mathbf{b}$ of the hexagonal cell. The overlapping spectra are consistent with the threefold rotational symmetry of the $P\bar{3}m1$ structure, which imposes $\varepsilon_{xx}(\omega)=\varepsilon_{yy}(\omega)$ and an isotropic in-plane linear optical response. Consequently, equivalent spectra are expected for any in-plane polarization direction, including those parallel to $\mathbf{a}$ and $\mathbf{b}$.

The absorptance exhibits a pronounced maximum of approximately \SI{25}{\percent} near \SI{1.4}{\electronvolt}, corresponding to a wavelength of about \SI{890}{\nano\metre} in the near-infrared region. Appreciable absorption persists across the visible spectrum, with values of approximately \SIrange{18}{21}{\percent}. Additional features occur near \SI{3.1}{\electronvolt} and \SI{4.8}{\electronvolt}, extending the optical response into the ultraviolet region. Above approximately \SI{5}{\electronvolt}, the absorptance progressively decreases, reaching about \SI{3}{\percent} at \SI{10}{\electronvolt}. The calculated spectrum therefore reveals a broad response encompassing the near-infrared, visible, and ultraviolet regions, with its strongest absorption occurring in the near-infrared.

Reflectance remains low throughout the investigated energy range, reaching a maximum of approximately \SI{2.1}{\percent} near the principal absorption peak. Transmittance follows the complementary trend, with a minimum of approximately \SI{73}{\percent} near \SI{1.4}{\electronvolt} and values around \SIrange{78}{82}{\percent} across the visible region. At higher photon energies, the decreasing absorptance is accompanied by increasing transmittance, which approaches \SI{97}{\percent} at \SI{10}{\electronvolt}. These results indicate that absorption accounts for most of the optical attenuation within the reported spectrum, whereas reflection contributes only a small fraction.

\subsection{Mechanical properties}

The in-plane elastic response was evaluated from the quadratic dependence of total energy on small applied strains. The threefold rotational symmetry of the trigonal structure constrains the in-plane linear-elastic response to an isotropic form, with two independent constants, $C_{11}$ and $C_{12}$, and $C_{66}=(C_{11}-C_{12})/2$. The calculated values, rounded to two decimal places, are $C_{11}=\SI{68.00}{\newton\per\metre}$, $C_{12}=\SI{40.68}{\newton\per\metre}$, and $C_{66}=\SI{13.66}{\newton\per\metre}$.

\begin{figure*}[t]
\centering
\includegraphics[width=\textwidth]{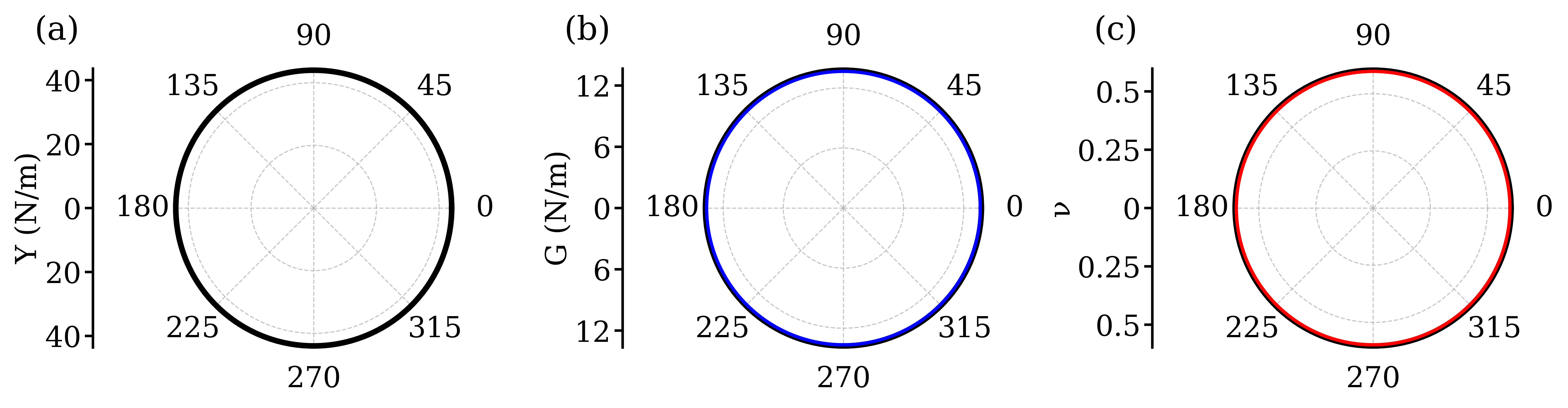}
\caption{Directional in-plane elastic properties of \cNb{}: (a) Young's modulus, $Y(\theta)$; (b) shear modulus, $G(\theta)$; and (c) Poisson's ratio, $\nu(\theta)$. Their constant values of \SI{43.66}{\newton\per\metre}, \SI{13.66}{\newton\per\metre}, and 0.598, respectively, indicate isotropic in-plane linear elasticity.}
\label{fig:elastic_polar}
\end{figure*}

The conditions $C_{11}>0$ and $C_{11}>|C_{12}|$ ensure positive definiteness of the isotropic 2D stiffness matrix~\cite{born1940stability}. Both are satisfied by \cNb{}. The eigenvalues of the in-plane stiffness matrix in engineering-shear Voigt notation are \SI{13.66}{\newton\per\metre}, \SI{27.32}{\newton\per\metre}, and \SI{108.68}{\newton\per\metre}. Their positive values confirm mechanical stability against infinitesimal in-plane strains.

The elastic moduli were obtained as $Y=(C_{11}^{2}-C_{12}^{2})/C_{11}$, $G=C_{66}$, and $K_{\mathrm{2D}}=(C_{11}+C_{12})/2$, giving \SI{43.66}{\newton\per\metre}, \SI{13.66}{\newton\per\metre}, and \SI{54.34}{\newton\per\metre}, respectively. Here, $K_{\mathrm{2D}}$ denotes the area bulk modulus. The ratio $K_{\mathrm{2D}}/G=3.978$ indicates greater resistance to uniform area change than to shear deformation. The relatively compliant response is consistent with a corrugated framework that may accommodate deformation through bond-angle changes and out-of-plane displacements, in addition to bond stretching.

The Poisson's ratio, $\nu=C_{12}/C_{11}=0.598$, indicates strong coupling between longitudinal and transverse strains. A longitudinal tensile strain of $1.00\%$ therefore corresponds to a transverse contraction of approximately $0.60\%$ in the linear regime. This value is admissible for isotropic 2D elasticity, where $-1<\nu<1$, although it exceeds the familiar upper limit of $0.5$ for isotropic 3D elasticity.

The isotropic stiffness tensor gives identical Voigt and Reuss area bulk and shear moduli, so $K_{\mathrm{V}}/K_{\mathrm{R}}=G_{\mathrm{V}}/G_{\mathrm{R}}=1$. The circular profiles in Fig.~\ref{fig:elastic_polar} likewise show no preferred in-plane direction for $Y$, $G$, or $\nu$. Thus, the out-of-plane corrugation and inequivalent \ce{Nb-Nb} distances coexist with isotropic in-plane linear elasticity.

For comparison, the reported maximum Young's moduli of goldene, copperene, and silverene are approximately \SI{73}{\newton\per\metre}, \SI{59}{\newton\per\metre}, and \SI{44}{\newton\per\metre}, respectively~\cite{10.1021/acsomega.5c01823}. The value for \cNb{} is close to that of silverene and lower than those of copperene and goldene. Its $C_{11}$ is comparable to the copperene value of \SI{71.11}{\newton\per\metre} and exceeds the silverene value of \SI{53.90}{\newton\per\metre}. However, the comparatively large $C_{12}$ lowers the relaxed uniaxial modulus through $Y=C_{11}-C_{12}^{2}/C_{11}$. Its shear constant is also smaller than the values reported for goldene (\SI{24.27}{\newton\per\metre}) and copperene (\SI{20.95}{\newton\per\metre}). Accordingly, \cNb{} combines appreciable resistance to area compression with greater compliance under uniaxial and shear loading.

\section{Conclusions}
\label{sec:conclusion}

First-principles calculations identify \cNb{} as a strongly cohesive, corrugated monoelemental 2D structure with a centrosymmetric $P\bar{3}m1$ symmetry, a hexagonal in-plane lattice, and a six-atom primitive cell. The nanosheet has a geometrical thickness of \SI{4.279}{\angstrom}, with short \ce{Nb-Nb} contacts ranging from \SI{2.685}{\angstrom} to \SI{2.887}{\angstrom}. Its cohesive energy of $-6.03~\mathrm{eV\,atom^{-1}}$ indicates strong binding relative to isolated atoms, while the absence of imaginary frequencies along the calculated phonon path supports harmonic dynamical structural stability.

The electronic structure exhibits multiband metallicity dominated by \ce{Nb} $4d$ states near the Fermi level. The calculated optical response is isotropic within the nanosheet plane, with a maximum absorptance of approximately \SI{25}{\percent} near \SI{1.4}{\electronvolt} in the near-infrared region, appreciable visible-light absorption, and low reflectance throughout the investigated energy range. The elastic constants satisfy the in-plane stability criteria and yield an isotropic linear-elastic response, with Young's and shear moduli of \SI{43.66}{\newton\per\metre} and \SI{13.66}{\newton\per\metre}, respectively, and a Poisson's ratio of 0.598. These results establish a combination of strong cohesion, mechanical compliance, and multiorbital metallicity, motivating further investigation of corrugated niobene for metallic and UV-responsive 2D systems.

\begin{acknowledgement}
This work was partially financed by the Brazilian Research Agencies CAPES and CNPq. 
The authors also thank the Distrito Federal Research Foundation (FAPDF) for financial 
and equipment support (Edital 04/2017 and Edital 09/2022). David L. Azevedo 
acknowledges financial support from the Mato Grosso Research Foundation (FAPEMAT) 
through Grant PRONEX CNPq/FAPEMAT 850109/2009, and from CNPq through Grant 
No. 315623/2021-7 (PQ-2). Kleuton A. L. Lima acknowledges financial support from 
the São Paulo Research Foundation (FAPESP) through Grant No. 2024/21870-8. 
This work used computational resources provided by the Centro Nacional de 
Processamento de Alto Desempenho em São Paulo (CENAPAD-SP).
D.S.G. acknowledges the Center for Computing in Engineering and Sciences at Unicamp for financial support through the FAPESP CEPID Grant(process number 2013/08293-7)and support from INEO/CNPq and FAPESP (grant 2025/27044-5).
\end{acknowledgement}

\bibliography{ref}
\end{document}